\documentclass[twocolumn,pre,aps,epsf]{revtex4-1}
\usepackage{graphicx}
\usepackage{amsmath,amssymb,amsbsy,amstext} 

\begin{document}

\title {Reply to ``Comment on `Phase transition in a network model of social balance with Glauber dynamics' ''}

\author{Pouya Manshour}
\email{manshour@pgu.ac.ir}
\affiliation{Physics Department, Persian Gulf University, Bushehr 75169, Iran}
\author{Afshin Montakhab}
\email{montakhab@shirazu.ac.ir}
\affiliation{Physics Department, Shiraz University, Shiraz 71454, Iran}

\begin{abstract}
Recently, we introduced [Physical Review E 100, 022303 (2019)] a
stochastic social balance model with Glauber dynamics which takes
into account the role of randomness in the individual's behavior.
One important finding of our study was a phase transition from a
balance state to an imbalance state as the randomness crosses a
critical value, which was shown to vanish in the thermodynamic
limit. In a recent similar study [K. Malarz and K. Ku\l akowskiy,
(2020), arXiv:2009.10136], it was shown that the critical
randomness tends to infinity as the system size diverges. This led
the authors to question our results. Here, we show that this
apparent inconsistency is the results of different definitions of
energy in each model. We also demonstrate that synchronous and
sequential updating rules can largely affect the results, in
contrast with the claims made by the aforementioned authors.

\end{abstract}

\maketitle

\section{ Introduction }
\label{intro}

In motivation psychology, Heider proposed the balance theory
\cite{Heider:1946} in $1946$. By considering the relationship
between three elements  which includes Person (P) and Other person
(O) with an object (X), known as the POX pattern, he postulated
that only balanced triads are stable. The POX is ``balanced" when
P and O are friends and they agree in their opinion of X. Balance
theory has been used to study the stability of a society,
where the individuals change their attitudes, beliefs, or
behaviors in order to reach a psychologically balance state. Much
research has been done to model and analyze different static and
dynamic aspects of the balance theory
\cite{Cartwright:1956aa,antal2005dynamics,marvel2009energy,Facchetti2011,marvel2011continuous}
in social networks where a set of elements or nodes like
countries, corporations, or people interact through different
types of connections or links, such as friendship (positive) or
hostility (negative). With this point of view, a social network is
balanced if it only includes balanced triads. Even though avoiding
distress and conflict is a natural tendency, this is not the case
when we deal with the real world. Generally speaking social
anomalies, as well as random activity of the individuals are
always present in social networks. Additionally, the presence of
jammed states \cite{antal2005dynamics,marvel2009energy} which can
trap the system in an imbalanced state posed a challenge for
achieving balance in a social network.

In a recent work \cite{Shojaei2019}, we introduced a dynamical
model which included an intrinsic randomness in the social agents
behavior. The model takes into account the possibility of both
increasing or decreasing the global tension via local transitions,
but on average, reduces the network's tension. In fact, using the
Glauber algorithm \cite{Glauber1963} with sequential update rules,
our dynamics is a finite temperature generalization of the
so-called \textit{Constrained Triad Dynamics} (CTD) model
\cite{antal2005dynamics}. Due to this feature, the dynamics is
able to escape out of jammed states and reach a final balanced
state, when the randomness is less than a critical value. We also
showed that such a critical value vanishes as the system size
diverges. In a recent work \citep{Malarz2020}, Malarz and Ku\l
akowskiy introduced a similar model with randomness as a heat-bath
algorithm with synchronous (parallel) update rule. They showed
that their system reaches a balanced state if the randomness is
less that a critical value. They found that this critical value
diverges for large system sizes, in contrast to our findings. In
this respect, they questioned the size dependency of the critical
randomness in our work. They also indicated that ``It seems
unlikely that the computational algorithm (the heat-bath vs the
Glauber dynamics) or links update order (synchronous vs
sequential) could be responsible for such a clear departure of the
results"\citep{Malarz2020}. In this paper, we show that this
inconsistency is rooted in the way energy is defined in each
model. Additionally, by using synchronous updating rule in our
model, we demonstrate that the final states of the system can be
quite different from the case with sequential updating rule.

\section{Models description}
\label{method} In this section, we explain both models in more
details. For simplicity, we name our model and the model
introduced by Malarz and Ku\l akowskiy, ``Model I" and ``Model
II", respectively.
\subsection{Model I}
In the model introduced in \cite{Shojaei2019}, we consider a fully
connected network of size $N$, and use a symmetric connectivity
matrix $s$, such that $s_{ij}=\pm1$. The positive sign represents
friendship, and the negative one represents hostility between two
arbitrary nodes $i$ and $j$. The total energy of the system is
defined as \cite{antal2005dynamics,marvel2009energy}:
\begin{equation}
U=\frac{1}{N_{tri}}\sum_{i>j>k} u_{ijk}
\label{glob_energy}
\end{equation}
where $u_{ijk}=-s_{ij}s_{jk}s_{ki}$, and the normalization factor
$N_{tri}=\binom{N}{3}$ is the total number of triads in the
network, so we have $-1\leq U \leq 1$. By this definition, we have
$u=-1$ and $+1$ for a balanced and an imbalanced triad,
respectively. Also, $U=-1$ is the balance condition for the
system, where all the triads are balanced. We update the system
sequentially, i.e., at every time step, we
flip a randomly chosen link with probability $p(t)$, defined as
\begin{equation}
p(t)=\frac{1}{1+e^{\beta \Delta U(t)}}
\label{p_glob}
\end{equation}
where $\beta$ is a control parameter which represents the inverse
of the randomness (temperature) in the system. Also, $\Delta U(t)$
indicates the change in the \textit{total} energy due to the
link-flipping in every time step $t$. This model can be considered
as the Glauber dynamics \cite{Glauber1963} used in simulations of
kinetic Ising model in contact with a heat-bath at temperature
$T=1/k\beta$, and thus, our model corresponds to a finite
temperature generalization of CTD model. It is important to note
that such dynamics provides a more realistic feature of creating
or reducing tension at any given time while \textit{on average}
reducing tension for positive finite $\beta$. Furthermore, since
it allows for increase in the energy of the system, it could
provide a natural mechanism to escape out of local minima, i.e.,
the jammed states. We showed that the final fate of the system is
a balanced state if the randomness is less than a critical value
($T<T_c$). We found that the system transitions from a balanced
into an imbalanced phase, when the randomness crosses its critical
value, and this critical randomness vanishes for infinite system
size. For $T<T_c$, we also found that for all initial positive
link densities of $\rho_0\le 1/2$, the system reaches a bipolar
state (two fully positive subgraphs that are joined by negative
links) with a final density of $\rho = 1/2$. On the other hand,
for the initial $\rho_0>1/2$, a paradise state (a fully positive
graph) with $\rho=1$ emerges. This eventual transition occurs at
the sharp value of $\rho_0=1/2$.

\subsection{Model II}
We now consider the model introduced in \citep{Malarz2020}. By
including the role of randomness in the individual's behavior, the
authors used a heat-bath algorithm to synchronously update the
state of the network. The time evolution of an arbitrary link
$s_{ij}(t)$ is given by
\begin{equation}
s_{ij}(t)=\left\{
                \begin{array}{lll}
                     +1 & \mbox{with probability} & p_{ij}(t) \\
                     -1 & \mbox{with probability} & 1-p_{ij}(t)
                \end{array}
         \right.
\end{equation}
where $p_{ij}(t)$ is given as
\begin{equation}
p_{ij}(t)=\frac{e^{\xi_{ij}(t)/T}}{e^{\xi_{ij}(t)/T}+e^{-\xi_{ij}(t)/T}}=\frac{1}{1+e^{-2\xi_{ij}(t)/T}}
\label{p_loc}
\end{equation}
and the local energy of each link is defined as
\begin{equation}
\xi_{ij}(t)=\sum_{k\neq i,j}s_{ik}(t)s_{kj}(t). \label{loc_energy}
\end{equation}
They showed that as long as the randomness is lower that a
critical value ($T<T_c$), the system reaches a paradise state
under any initial conditions except for $\rho_0\approx 1/2$, at
which the bipolar state with $\rho=1/2$ emerges. They also found
that this critical randomness diverges for infinite system size.
Their main conclusion is that this finding is in conflict with our
results.

In the following, we will demonstrate that the underlying
mechanism and definition used in Model II are different from those
of Model I, and thus one can expect different results, not only in
the behavior of the critical temperature but also in the eventual
states of the system. Clearly, the most important parameter that
governs the dynamics in both models is the probability $p(t)$
(Eq.~\ref{p_glob}) or $p_{ij}(t)$ (Eq.~\ref{p_loc}).  For example,
if it approaches the $1/2$ value, each link's sign is changed
completely randomly. In this respect, the final state of the
system is an imbalanced state with nearly equal number of positive
and negative links, so that $\rho=1/2$. Another important quantity
is the energy which is used to calculate the probabilities. The
main difference between the two models can be seen in their
different definition of energy.

Since each link is connected to $N-2$ triads,  changing the sign
of a given link affects $N-2$ terms in the total energy,
Eq.~\ref{glob_energy} in Model I. Thus, based on the normalization
factor ,$N_{tri}=\binom{N}{3}=N(N-1)(N-2)/3!$, the energy
difference $\Delta U(t)$ in the flipping probability of
Eq.~\ref{p_glob} scales roughly as
\begin{equation}
|\Delta U(t)|\sim \frac{N-2}{N(N-1)(N-2)}\sim \frac{1}{N(N-1)}
\end{equation}
In other words, $|\Delta U|\to 0$ as $N\to \infty$. This indicates
that for large systems, $p\to 1/2$, pushing the system into a
random imbalanced configuration, unless the randomness also
vanishes. Therefore, the phase transition from an imbalanced state
into a balanced state occurs at $\beta\to \infty$ for infinite
system size, indicating that $T_c(N\rightarrow\infty)\rightarrow
0$.

In Model II, nearly the same algorithm has been applied but using
a different definition for the energy. Model II assigns a (local)
energy $\xi_{ij}$ to each link $s_{ij}$ as a sum over all $N-2$
products of link pairs ($s_{ik}s_{kj}$) neighboring the link
$s_{ij}$ (Eq.~\ref{loc_energy}). Therefore, one can simply find
that this energy diverges with system size as $|\xi_{ij}|\sim
N-2$, which means that only for large enough temperatures, can a
large system lead to an imbalanced random state ($p_{ij}\to 1/2$).
In other words, the critical temperature increases with increasing
system size $N$ as shown in Ref. \citep{Malarz2020}. Therefore, in Model
II, the balanced state is extremely robust to random fluctuations
while in Model I, it is extremely vulnerable. This is due to how
the energy landscape is effected by the system size in a complete
network where all individuals interact with all other ones. In the
thermodynamics limit, the typical size of energy barriers diverge
in Model II, while they vanish in Model I.

\section{Synchronous versus sequential updating}

\begin{figure}[t]
\begin{center}
\includegraphics[scale=0.6]{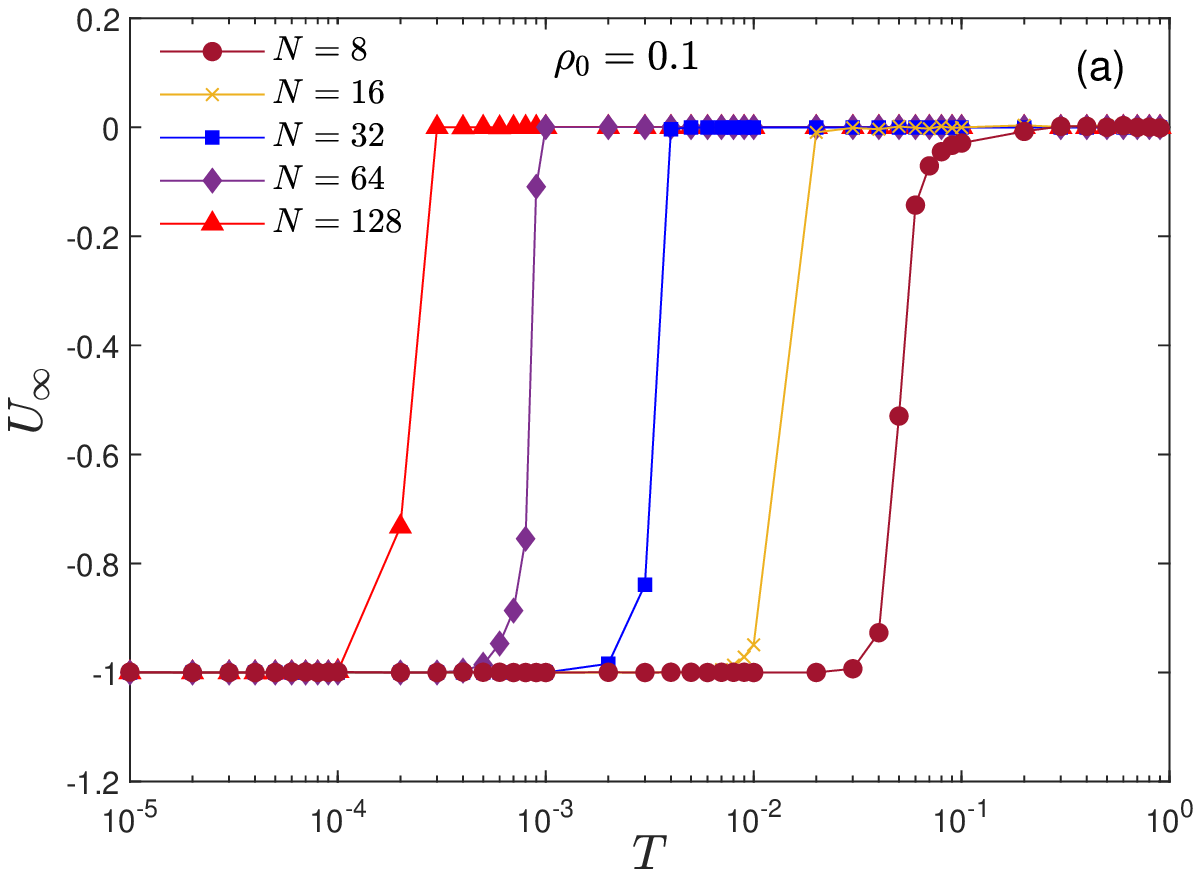}
\includegraphics[scale=0.6]{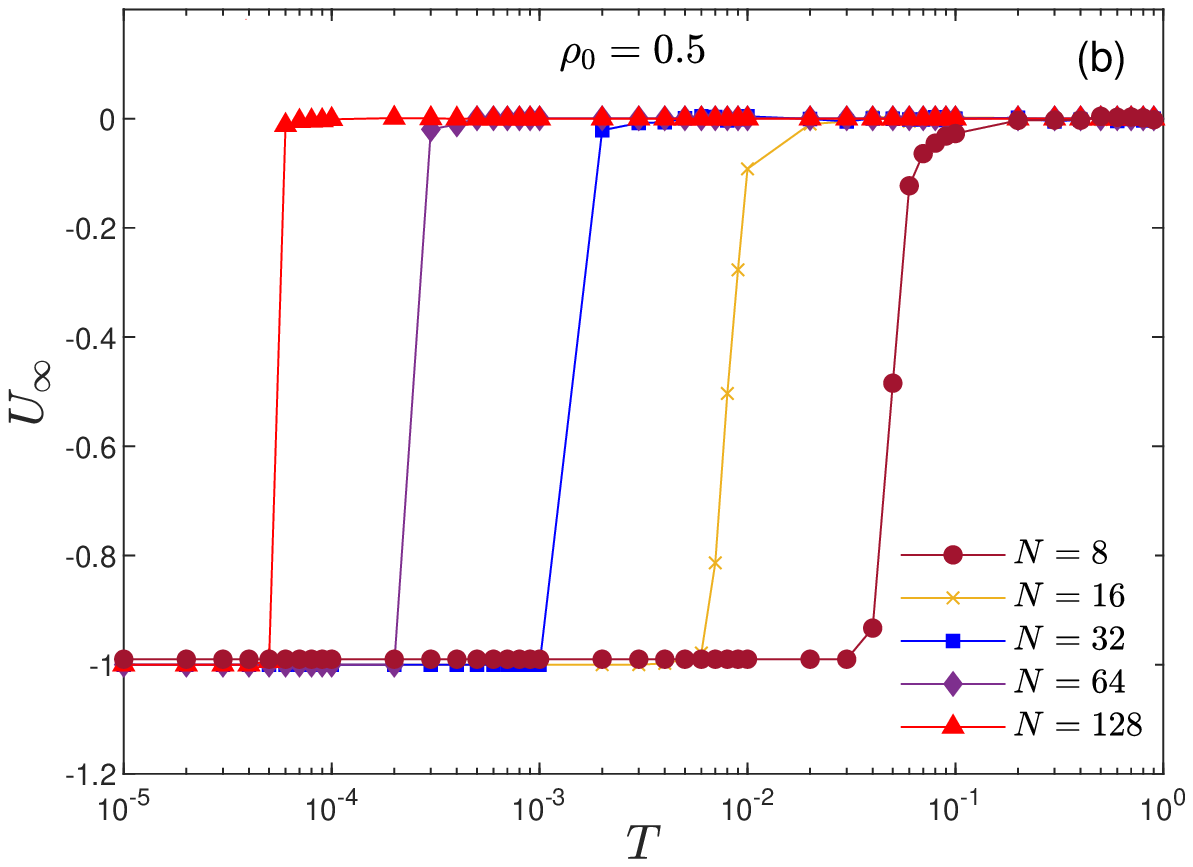}
\caption{The temperature ($T$) dependence of long time behavior of
the total energy, $U_\infty$, in Model I with parallel update for
(a) $\rho_0=0.1$ and (b) $\rho_0=0.5$. The system transitions from
a balance phase into an imbalance one, as the temperature crosses
a critical value $T_c$ which vanishes for arbitrary large
networks.} \label{fig1}
\end{center}
\end{figure}

\begin{figure}[t]
\begin{center}
\includegraphics[scale=0.6]{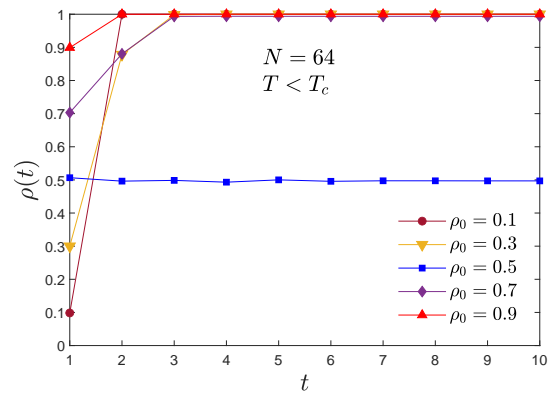}
\caption{The time evolution of the density of positive links,
$\rho(t)$, with different initial conditions of $\rho_0=0.1$,
$0.3$, $0.5$, $0.7$, and $0.9$, for $T=10^{-4}$ ($<T_c$) in Model
I with synchronous update rule. The network size for all plots is
$N=64$.} \label{fig2}
\end{center}
\end{figure}

Another important distinction between Model I and Model II is the
way their dynamics is implemented numerically, i.e. synchronous vs
sequential updating. In this section, we simulate Model I using
synchronous (parallel) update rule instead of the originally used
sequential dynamics. Therefore, given that at time
$t$ energy is given by $U(t)$, for time $t+1$, we update the state
of all links synchronously as follows: each link is flipped and
then we calculate the total energy of the system, $U$, due to this
link's flipping. Next, by calculating $\Delta U=U-U(t)$, with
probability $p$ defined in Eq.~\ref{p_glob}, this flipping is
accepted or otherwise the link sign does not change. Then, we
bring the system back to its previous state, and repeat the above
mentioned updating rule for the next link. This is done for each
individual link and finally all such resulting states are
implemented in one time step, i.e. $t+1$.

Using such parallel update rule, we calculate the final total
energy of the system, $U_\infty$, for different values of the
temperature, $T$, and also for various initial positive link
densities, $\rho_0$. We represent in Figs.~\ref{fig1}(a) and
\ref{fig1}(b), the values of $U_\infty$ versus $T$ for two initial
positive link densities of $\rho_0=0.1$ and $\rho_0=0.5$, for
different system sizes of $N=8$, $16$, $32$, $64$, and $128$. As
can be seen, the system transitions from a balanced state
($U_\infty =-1$) into an imbalanced one ($U_\infty=0$), as it
crosses a critical value. This critical value decreases as the
system size increases, so that $T_c \to 0$ for large $N$. This
indicates that the difference in the update rule is not
responsible for the different scaling behavior of the critical
temperature in Model I and Model II. Indeed, the difference in the
behavior of the critical temperature is due to the scaling of the
energy landscape.

One can see from Fig.1 that the critical temperature also depends
on the initial link density $\rho_0$. Another important difference
between Model I (sequential) and Model II (synchronous) is their
balanced states dependance on the initial link density. In Model
I, we found that the final state is a paradise state ($\rho=1$)
for all initial link densities greater that half, and is a bipolar
state ($\rho=1/2$) for all initial link densities smaller than
half.  On the other hand, Model II was shown to lead to a paradise
state for all initial link densities except for values close to
half, i.e. $\rho_0\approx 1/2$ where it resulted in a bipolar
state. Since both balanced states are the attractor of the
stochastic dynamics, one can suspect that such discrepancy is due
to the update rules implemented.  We therefore study Model I with
synchronous dynamics in order to find the dependence of the final
state on the initial conditions. The results are shown in Fig.2.
As can be seen, one achieves results similar to Model II once one
uses synchronous update for Model I.  Our results seem to indicate
that sequential update allows smooth flows in state space in order
to discover the nearby attractors, while parallel update favors
transitions to the paradise state, as it causes large transitions
in the state space. It is important to note that previous studies
have also found significant difference between various update
rules within a given model \cite{Rolf1998,Schmoltzi1995,Shu2016}.

In conclusion, we have shown that the claimed discrepancy in the
critical behavior of the temperature in Ref.\cite{Shojaei2019}
(Model I) compared to that in Ref.\cite{Malarz2020} (Model II) can
be understood in the scaling property of the energy (landscape) as
a function of system size.  Furthermore, we have shown that such
models behave distinctly differently when updated synchronously
vs. sequentially.  In fact, the final balanced state strongly
depends on how one updates the dynamics.

\bibliographystyle{apsrev4-1}

\end{document}